\documentclass[reprint,superscriptaddress,aps,prx]{revtex4-2}

\usepackage{graphicx}
\usepackage{circledsteps}
\usepackage{amssymb}
\usepackage{amsmath}
\usepackage{epsfig}
\usepackage{chemarr}
\usepackage{url}
\usepackage{natbib}
\usepackage{dcolumn}
\usepackage{bm}
\usepackage{xcolor}
\usepackage{braket}

\usepackage[
colorlinks=true,
linkcolor=blue,
citecolor=blue,
urlcolor=blue
]{hyperref}

\newcommand{\gex}{\gamma_{\mathrm{ex}}}
\newcommand{\gin}{\gamma_{\mathrm{in}}}

\newcommand{\Ploc}{P_{\mathrm{loc}}}
\newcommand{\floc}{f_{\mathrm{loc}}}
\newcommand{\Rcoh}{\mathcal{R}}

\begin{document}

\title{An interdimensional funnel enhances topologically protected oscillations in high-dimensional stochastic systems}

\author{Varsha Traynor}
\address{Department of Physics and Astronomy, University College London, London WC1E 6BT, United Kingdom}
\author{Jaime Agudo-Canalejo}
\email{j.agudo-canalejo@ucl.ac.uk}
\address{Department of Physics and Astronomy, University College London, London WC1E 6BT, United Kingdom}

\begin{abstract}
Biochemical systems such as protein complexes often occupy configuration spaces whose dimensionality grows with the number of molecular components. Although usually viewed as an obstacle to coherent dynamics, here we show that high dimensionality can enhance topologically protected oscillations. In the topological regime, a stochastic system with a $D$-dimensional configuration space develops a steady-state current confined to a one-dimensional edge cycle. The intervening boundary faces between the $D$-dimensional bulk and the one-dimensional cycle form an interdimensional funnel, sequentially pushing the system towards lower dimensional faces, down to the one-dimensional cycle. As a consequence, increasing $D$ enhances the funnel, leading to improved one-dimensional localization and oscillatory coherence. A quantized biorthogonal Zak phase identifies the transition to the topological regime, which additionally exhibits a highly structured non-Hermitian skin effect. In contrast with previous realizations of topological states in stochastic systems, the phenomena we uncover have no counterpart in quantum condensed matter or active matter systems.
\end{abstract}

\maketitle

\section{Introduction}

Many biological machines are assembled from repeated subunits \cite{marsh2015structure}. Proteins that form ring complexes such as the cyanobacterial clock protein KaiC \cite{SwanEtAl2018} and the  memory- and learning-associated protein CaMKII \cite{BhattacharyyaEtAl2020} can therefore occupy combinatorially large configuration spaces: if each of $D$ subunits has $N$ possible modification or conformational states, the complex has $N^D$ configurations. Cooperative equilibrium models reduce this high dimensionality by assigning interactions between subunits \cite{Monod1965,Koshland1966}. Yet many molecular cycles are driven by nucleotide hydrolysis and violate detailed balance \cite{Schnakenberg1976,Seifert2012}. As a consequence, their collective behavior is set not only by an energy landscape but also by directed probability currents through configuration space \cite{tu2008nonequilibrium,mahdavi2024flexibility,dgw7-s8gl}. The Kai system, whose ordered phosphorylation cycle persists in vitro, provides a prominent example \cite{Nakajima2005,Rust2007,VanZon2007}.

Topology offers a route to robust dimensionality reduction independent of the details of individual transitions. Ideas adapted from topological quantum matter have produced localized modes and protected currents in classical, active, and stochastic systems \cite{Shankar2022,agudo2025topological,sone2026hermitian,Murugan2017,Dasbiswas2018,sawada2024role,Knebel2020,Tang2021,yoshida2022non,edwards2026robusttopologicallyprotectededge,kuroda2026designingtopologicaledgecurrents}. In a two-dimensional stochastic model, nonequilibrium futile cycles in the bulk were shown to generate a chiral current at the boundary of configuration space \cite{Tang2021}. While this mechanism suggested a way to achieve robust biochemical cycles, its reliance on a two-dimensional configuration space strongly restricted its applicability to generic biological protein complexes with an arbitrary-dimensional configuration space. In particular, the boundary of a high-dimensional space is itself high-dimensional, and it is therefore unclear whether a high-dimensional system can robustly collapse onto a useful one-dimensional trajectory.

Here, we show that stochastic edge currents exist in any arbitrary dimension, and reveal a striking geometric effect absent in two dimensions. The many intermediate-dimensional boundaries (i.e., the $d$-dimensional faces with $1<d<D$)  between the $D$-dimensional bulk and the one-dimensional cycle do not dilute the edge response. Instead, they act as a directed cascade that funnels probability toward the cycle. Increasing dimension can thus make the localized current both larger and more coherent. We establish this result from stochastic trajectories and steady-state distributions, a geometric argument for the size of the funnel, temporal and spectral measures of coherence, as well as through a bulk topological invariant and a characterization of the associated non-Hermitian skin effect.

\section{High-dimensional futile cycles}

\begin{figure*}
\centering
\includegraphics[width=\linewidth]{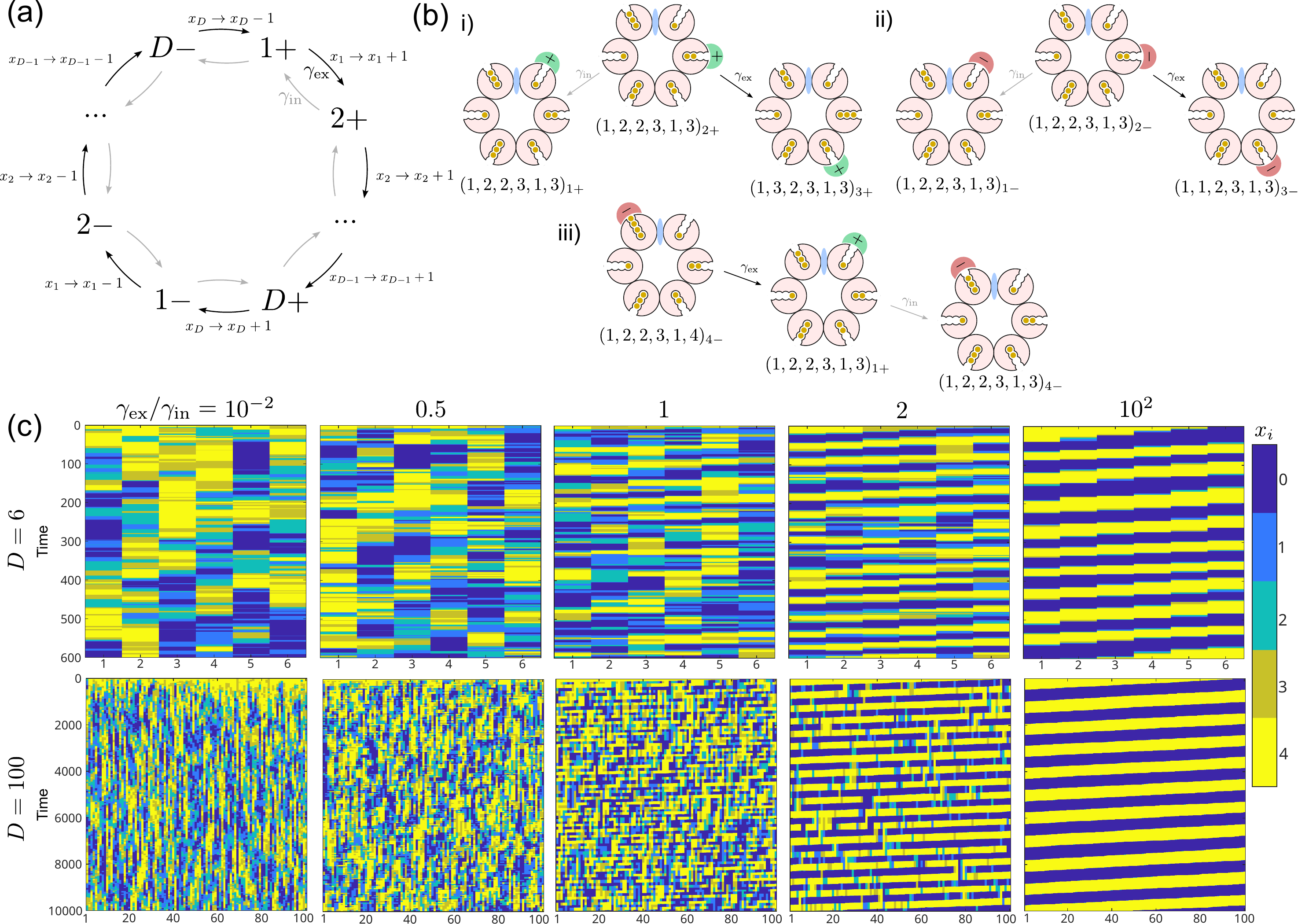}
\caption{\textbf{Futile cycles enable a global edge cycle in arbitrary dimensions.}
(a) Abstract depiction of the stochastic dynamics. The $2D$ internal states form a ring. External transitions
(black, rate $\gex$) change the internal state clockwise while also changing the external state, incrementing or decrementing $x_n$ when leaving state $n+$ or $n-$, respectively. Internal transitions
(gray, rate $\gin$) change the internal state anticlockwise without affecting the external state.  (b) Molecular implementation on a protein ring of $D$ phosphorylatable
subunits. The external state corresponds to the phosphorylation state of all subunits.  ({i},{ii}) The internal state $n+$ or $n-$ indicates an enzyme bound to the $n$th subunit and primed to phosphorylate or dephosphorylate, respectively. ({iii}) The enzyme switches between the $+$ and $-$ states when it crosses over the regulatory site in blue. (c) Stochastic trajectories of all $D$ coordinates for several values of $D$ (rows) and $\gex/\gin$ (columns), for fixed $N=5$. Color denotes the coordinate
value from 0 to 4. Below $\gex/\gin=1$, the coordinates fluctuate
incoherently. Above this value, ordered diagonal bands reveal a global edge cycle,
which becomes more regular with increasing $D$. Time is reported in
units of $\gin^{-1}+\gex^{-1}$.}
\label{fig:futile}
\end{figure*}

We consider a continuous-time Markov process whose microscopic state is determined by $(x_1,\ldots,x_D)_s$. The `external' state $(x_1,\ldots,x_D)$ lies on a bounded hypercubic lattice, with coordinates $x_n\in\{0,\ldots,N-1\}$, while $s$ is one of $2D$ `internal' states labelled $1+,2+,\ldots,D+,1-,2-,\ldots,D-$. The internal states can be arranged in a ring [Fig.~\ref{fig:futile}(a)]. An external transition, occurring with rate $\gex$, advances clockwise around this ring and simultaneously changes one external coordinate: leaving $n+$ increments $x_n$ by one, whereas leaving $n-$ decrements it. The external transition is absent when the corresponding coordinate change would leave the hypercube, i.e.~there is no external transition starting from $n+$ ($n-$) if $x_n=N-1$ ($x_n=0$). An internal transition, occurring with rate $\gin$, advances one step anticlockwise without changing the external state. Note that this stochastic dynamics defines two types of local `futile' cycles: a clockwise external cycle of $2D$ consecutive external transitions increments and subsequently decrements every coordinate by one, returning to the original configuration; an anticlockwise internal cycle of $2D$ consecutive internal transitions similarly returns to the initial configuration while retaining the same coordinate throughout the cycle.

The network of stochastic transitions thus defined forms a hypercubic lattice with $N^D$ unit cells, each containing $2D$ sites, for a total of $N_{\mathrm{tot}}=2D N^D$ microscopic states. For $\gex,\gin>0$, the corresponding directed graph is strongly connected, implying that the Markov process is irreducible, and thus that the system has a unique steady state and is ergodic. In the case $D=2$, the model reduces to the square lattice studied in Ref.~\citenum{Tang2021}. Examples of the $D=2$ square lattice as well as the $D=3$ cubic lattice are shown in Supplementary Fig.~S1.

This abstract description can be mapped to various concrete microscopic dynamics. An example of a natural molecular realization is depicted in Fig.~\ref{fig:futile}(b). Consider a ring-like protein complex of $D$ phosphorylatable subunits and an enzyme in either a $+$ or $-$ state that is bound at subunit $n$. Each subunit can be phosphorylated up to $N-1$ times. In the $+$ ($-$) state, the enzyme can phosphorylate (dephosphorylate) subunit $n$ and advance to subunit $n+1$ at a rate $\gex$, or it can move back to subunit $n-1$ without effecting any changes on subunit $n$ at a rate $\gin$. Importantly, if the subunit is fully phosphorylated (dephosphorylated), then an enzyme in state $+$ ($-$) can only undergo the $\gin$ transition. Lastly, a regulatory motif causes a switch between the $+$ and $-$ states of the enzyme when it moves between subunit $D$ and subunit $1$ in either direction. Beyond this specific mapping, given here as an example, the external coordinate $x_n$ could represent other postranslational modifications, occupancies, or conformational states, while the internal state $s$ could represent tense or relaxed states of the relevant subunits or arise from the binding of more than one enzyme to the protein complex.

\section{Global cycles emerge from local cycles}

Naively, one may expect that the typical dynamics for $\gex\gg\gin$ ($\gex \ll \gin$) would simply consist of the system undergoing external (internal) futile cycles at the local level, punctuated by occasional internal (external) transitions that lead to effective diffusion across the whole lattice over time. This is indeed what is observed for stochastic simulations of the model  in an infinite lattice ($N\to\infty$) or for modified dynamics with periodic boundary conditions (i.e.~allowing transitions between $x_n=0$ and $x_n=N-1$). For a finite lattice, on the other hand, stochastic simulations (performed using the Gillespie method \cite{gillespie1976general}) show that, while random diffusion is indeed observed for $\gex/\gin \ll 1$, the system's behavior changes radically as $\gex/\gin \approx 1$, and global (system-spanning) cycles are observed instead as $\gex/\gin \gg 1$ [Fig.~\ref{fig:futile}(c)]. More specifically, in contrast to the local external futile cycles that the model imposes microscopically, which in the molecular implementation of Fig.~\ref{fig:futile}(b)  should induce a single phosphorylation event for each subunit, progressing in clockwise order, followed by a single dephosphorylation event in the same order; we instead observe complete phosphorylation of each subunit, progressing in anticlockwise order, followed by complete dephosphorylation in the same order.

This behavior is analogous to that found for the two-dimensional ($D=2$) case in Ref.~\citenum{Tang2021}. In this case, as well as in the $D=3$ case, the emergence of global cycles as $\gex/\gin \gg 1$ can be understood from visual inspection of the full lattice (Supplementary Fig.~S1). Interestingly, the stochastic simulations of Fig.~\ref{fig:futile}(c) suggest that, with increasing $D$, the cycling behavior becomes more coherent, and the transition at $\gex/\gin = 1$ becomes sharper. At $D=100$, the diagonal bands representing travelling waves of phosphorylation and dephosphorylation become strikingly regular. This ordered sequence is the dynamical signature of dimensional reduction from $D$ fluctuating coordinates to a one-dimensional directed cycle. In the following, we will refer to $\gex/\gin>1$ as the topological regime, although the topological nature of the transition will only be demonstrated in the final section.

\section{High dimensionality enhances edge localization}

\begin{figure*}
\centering
\includegraphics[width=\linewidth]{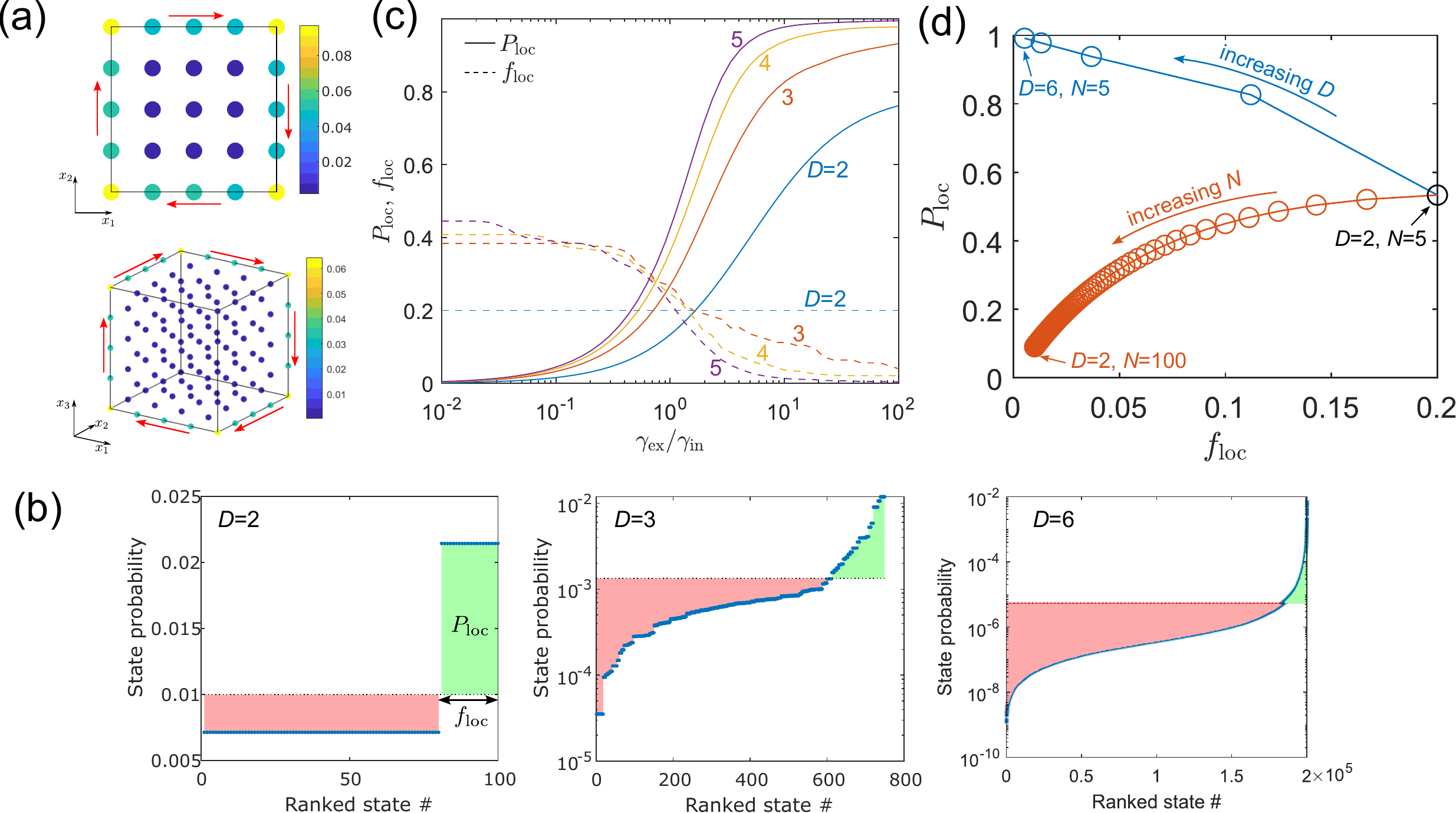}
\caption{\textbf{Higher dimensionality induces stronger localization onto a smaller fraction of states.}
(a) Steady-state probability of external states (summed over internal states) for
$D=2$ (top) and $D=3$ (bottom), with arrows indicating the direction of the steady-state edge current (global cycle). 
 (b) Rank-ordered steady-state probabilities for several values of $D$. Dotted lines mark the
value corresponding to a uniform distribution; green and red areas indicate excess and deficit,
respectively, from which the localization measures $\Ploc$ and $\floc$ are defined. (c) Total variation distance $\Ploc$ (solid) and
fraction of localized states $\floc$ (dashed) versus $\gex/\gin$ for
several values of $D$. (d) Parametric plot of $\Ploc$ against
$\floc$, representing an increase in $D$ from 2 to 6 at fixed 
$N=5$ (blue circles); or an increase in $N$ from 5 to 100 at fixed $D=2$ (red circles).
In (a), (b), and (c), $N=5$. In (a) and (b), $\gex/\gin=10^2$. In (d), $\gex/\gin=10$.}
\label{fig:localization}
\end{figure*}

To understand the effect of dimensionality, we calculate the steady state probability $p^{\,\mathrm{ss}}_\alpha$ for all states $\alpha=1,\ldots,N_\mathrm{tot}$ from the null right eigenvector of the transition matrix (Methods). For $D=2$ and $D=3$ in the topological regime, probability is depleted in the interior and accumulates on a distinguished one-dimensional path along the boundary, i.e.~the global cycle [Fig.~\ref{fig:localization}(a)]. Importantly, while the $D=3$ lattice contains two-dimensional faces, steady state probability does not accumulate in them and instead localizes to the global cycle composed of six of its twelve one-dimensional edges. The associated steady-state current (red arrows) follows this path with the opposite chirality to the external futile cycle. Generally, the global cycle includes $2D(N-1)$ external states arranged over $2D$ edges of the $D$-dimensional hypercube. Note that a $D$-dimensional hypercube has $D2^{D-1}$ edges in total, and thus only a fraction $2^{2-D}$ of them (which becomes increasingly small with increasing $D$) participates in the global cycle.

Ranked microscopic state probabilities reveal how strongly this localization depends on dimension [Fig.~\ref{fig:localization}(b)]. For $D=2$, the steady-state distribution separates into two constant and closely spaced levels. For $D=3$ and especially $D=6$, the distribution instead spans orders of magnitude. A small tail of states lies far above the uniform value $p_{\mathrm{u}}=1/N_{\mathrm{tot}}$, whereas most states are depleted. We quantify the excess by the total variation distance
\begin{equation}
\Ploc=\frac{1}{2}\sum_{\alpha=1}^{N_{\mathrm{tot}}}
\left|p^{\,\mathrm{ss}}_\alpha-\frac{1}{N_{\mathrm{tot}}}\right|,
\label{eq:ploc}
\end{equation}
which ranges from 0 for a uniform steady state to $1-1/N_\mathrm{tot}$ in the limit of complete accumulation in a single state, and represents the total probability displaced from underpopulated to overpopulated states (relative to the uniform value). We additionally define $\floc$ as the fraction of states whose probability exceeds the uniform value.

Both observables change near $\gex/\gin=1$ [Fig.~\ref{fig:localization}(c)]. Above the transition, $\Ploc$ rises and $\floc$ falls. The change becomes progressively sharper from $D=2$ to $D=5$, indicating that higher dimension concentrates a larger fraction of the steady-state probability into a smaller fraction of state space. Importantly, the trend is not a generic consequence of increasing the total number of states. In particular, increasing the dimensionality $D$ of the system at fixed size $N$, or increasing $N$ at fixed $D$ have radically different effects [Fig.~\ref{fig:localization}(d)]. Starting from $D=2$ and $N=5$, increasing $N$ at fixed $D=2$ leads to weaker localization ($\Ploc\to 0$) into an increasingly smaller number of states ($\floc\to 0$). Increasing $D$ at fixed $N=5$, on the other hand, leads to stronger localization ($\Ploc\to 1$) into an increasingly smaller number of states ($\floc\to 0$).

\section{The interdimensional funnel}

\begin{figure*}
\centering
\includegraphics[width=\linewidth]{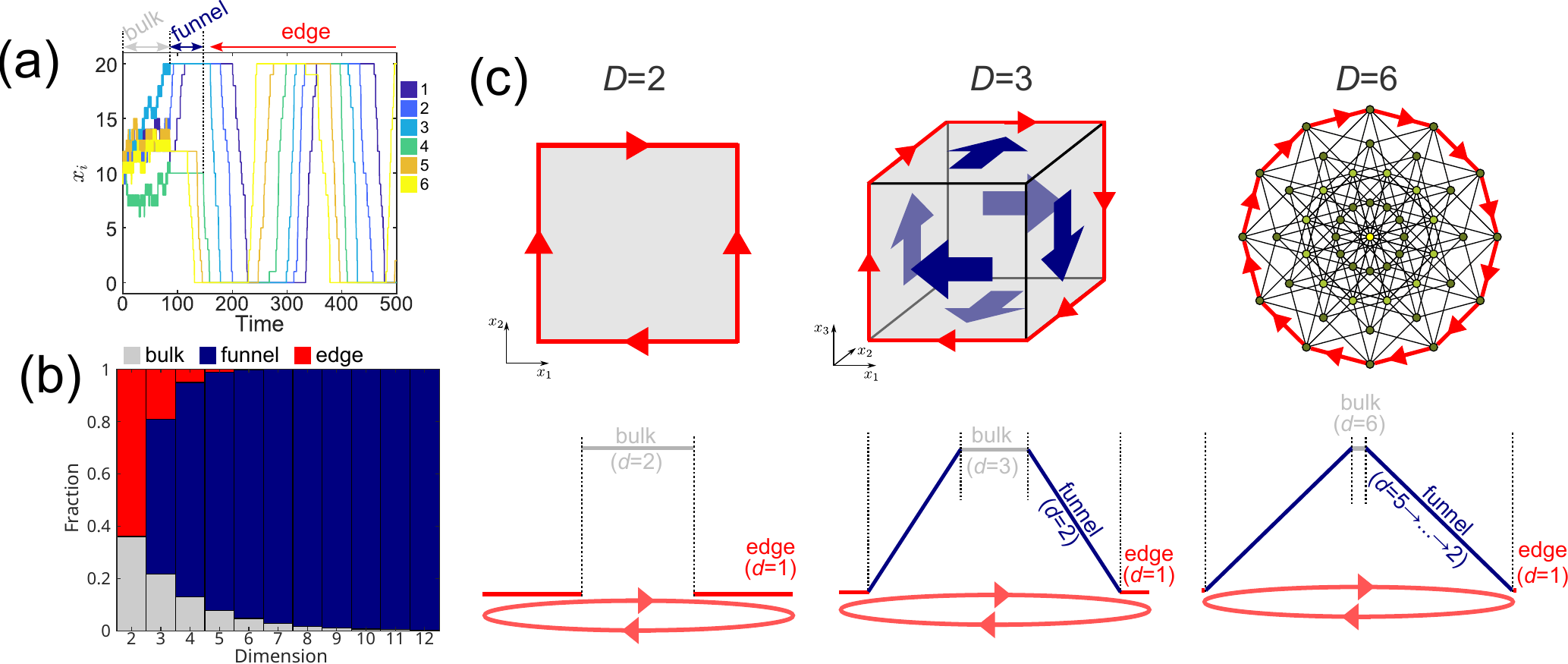}
\caption{\textbf{Boundaries funnel trajectories to the global cycle.}
(a) A stochastic trajectory (with $D=6$, $N=21$, and $\gex/\gin=2$) starting from the center of the bulk shows that the dynamics can be separated into three intervals: bulk diffusion
(gray), biased dynamics along intermediate-dimensional boundaries (funnel, blue), and oscillatory dynamics on the one-dimensional edge (red). The six coordinates are
colored as indicated. (b) Fractions of bulk, funnel, and edge cycle
configurations versus $D$ for $N=5$, calculated from Eqs.~\ref{eq:fractions}--\ref{eq:fractions2}.
(c) Geometric interpretation of the funneling effect for $D=2$, $D=3$, and $D=6$. Top row: depiction in real space, with the edge cycle in red. Bottom row: schematic depiction with the width of the bulk, funnel and edge represented to scale relative to the corresponding fractions in (b). In $D=2$, the bulk feeds the edge cycle directly. In $D=3$, two-dimensional faces direct (blue arrows) trajectories  toward  the edge cycle. In $D=6$, boundaries of increasingly lower dimension funnel trajectories toward the
edge cycle. The network at the top is a two-dimensional projection of the six-dimensional hypercube.}
\label{fig:funnel}
\end{figure*}

Why does increasing dimensionality enhance edge state localization, despite the edge state becoming an increasingly small fraction of the configuration space? An important observation is that, except in the special case $D=2$, a trajectory entering the boundary from the bulk does not generically reach the one-dimensional cycle. It first reaches a configuration where only one of its coordinates is at a boundary value ($x_n=0$ or $N-1$) while the others are not ($0<x_m<N-1$ for $m\neq n$), i.e.~a $(D-1)$-dimensional face or $(D-1)$-face. A stochastic simulation demonstrating this first contact for a system with $D=6$ and $N=21$ in the topological regime is shown in Fig.~\ref{fig:funnel}(a). In the bulk, the system undergoes random diffusion as expected. As soon as one coordinate reaches a boundary value (here $x_3=20$), however, we observe that all other coordinates are sequentially driven to a boundary value as well, in decreasing order (here, $x_2,x_1$ are first driven up to 20, then $x_6,x_5$ are driven down to 0). Once all but one coordinate (here, $x_4$) has been driven to a boundary value, the system has reached the one-dimensional edge corresponding to the global cycle, resulting in oscillatory dynamics.

In geometric terms, these dynamics imply that, as soon as the system reaches from the $D$-dimensional bulk into a $(D-1)$-face via random diffusion, it experiences a drift driving it into a $(D-2)$-face, and so on until it reaches a 2-face, which finally drives the system into the one-dimensional global cycle. We refer to these driven dynamics at the intervening boundary $d$-faces with $1<d<D$ as the interdimensional funnel. The size of this funnel can be estimated by a simple counting argument, ignoring internal states for simplicity. There are $(N-2)^D$ configurations in the bulk (i.e.~with no coordinate at a boundary) and $N^D$ configurations in total, so the fraction of bulk states is
\begin{equation}
f_{\mathrm{bulk}}=\left(\frac{N-2}{N}\right)^D=\left(1-\frac{2}{N}\right)^D.
\label{eq:fractions}
\end{equation}
The one-dimensional global cycle contains $2D(N-1)$ configurations, or a fraction
\begin{equation}
f_{\mathrm{edge}}=\frac{2D(N-1)}{N^D}.
\end{equation}
We define the remaining intervening boundary configurations as the funnel,
\begin{equation}
f_{\mathrm{funnel}}=1-f_{\mathrm{bulk}}-f_{\mathrm{edge}}.
\label{eq:fractions2}
\end{equation}
For the special case $D=2$, there are no intervening dimensions between the bulk and the cycle and $f_{\mathrm{funnel}}=0$. Importantly, we find that as $D$ increases at fixed $N>2$, both $f_{\mathrm{bulk}}$ and $f_{\mathrm{edge}}$ vanish, so that $f_{\mathrm{funnel}}\rightarrow1$, i.e.~the system becomes ``all funnel'' [Fig.~\ref{fig:funnel}(b)]. This is an example of the concentration of measure phenomenon: most of the volume of a high-dimensional object is concentrated near its surface.

This counting argument explains why high dimensionality improves dimensional reduction rather than obstructing it, even if the bulk dynamics are diffusive in all dimensions [Fig.~\ref{fig:funnel}(c)]. In two dimensions, the system enters the edge cycle directly from the bulk. In three dimensions, the system first encounters a two-dimensional face which then directs it toward the cycle (blue arrows in Fig.~\ref{fig:funnel}(c), top-center). In arbitrary dimension $D>2$, the system encounters a $(D-1)$-face, and a sequence of increasingly lower-dimensional faces funnel it toward the cycle. Crucially, at large $D$, the size of the bulk is negligible and the funnel occupies almost all configuration space, yet its dynamics are biased toward the one-dimensional cycle when $\gex>\gin$.

\section{High dimensionality increases cycle coherence}

\begin{figure}
\centering
\includegraphics[width=\linewidth]{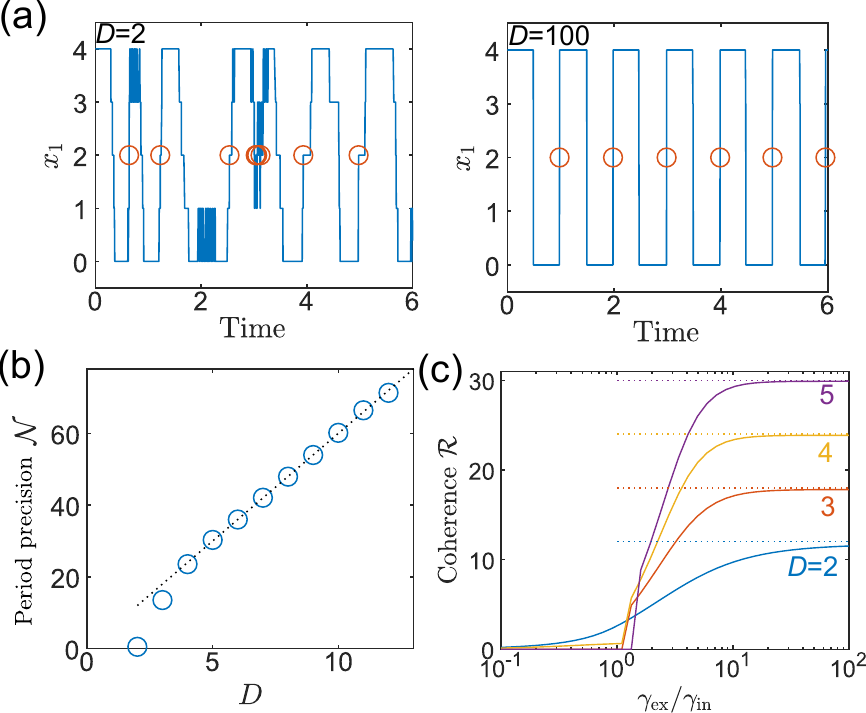}
\caption{\textbf{Higher dimensionality increases oscillatory coherence.}
(a) Representative stochastic trajectories of coordinate $x_1$  for $D=2$ and $D=100$, with $N=5$ and $\gex/\gin=10$. Time is in units of $2ND\gin^{-1}$, and circles mark
upwards crossings of $x_1=2$, defining an oscillation period. (b) Period precision of the oscillations
versus dimensionality $D$ for $N=3$ and $\gex/\gin=10^2$. (c) Spectral coherence
versus $\gex/\gin$ for $N=3$ and several values of $D$. In (b) and (c), the dotted lines correspond to $\mathcal{N}\approx\Rcoh\approx 2DN$ valid at large $\gex/\gin$.}
\label{fig:coherence}
\end{figure}

We next ask whether high dimensionality also makes the system a better oscillator. Figure~\ref{fig:coherence}(a) compares the dynamics of a representative coordinate ($x_1$) for $D=2$ and $D=100$ at $N=5$ and $\gex/\gin=10$. At $D=2$, repeated upward crossings of the midpoint (red circles) are unequally spaced and interrupted by rapid recrossings or incomplete cycles. At $D=100$, the same observable approaches a square wave with one clean and equally spaced midpoint crossing per cycle.

From successive upward crossings we measure the period $T$ and its variance, from which we can define the period precision
\begin{equation}
\mathcal{N}=\frac{\langle T\rangle^2}{\operatorname{var}(T)}.
\label{eq:fano}
\end{equation}
which measures the coherence of the oscillations. We find that it grows approximately linearly with $D$ at fixed $N$ and fixed $\gex/\gin \gg 1$ (Fig.~\ref{fig:coherence}(b), where $N=3$ and $\gex/\gin = 10^2$ are used). 

An alternative measure of coherence follows from the spectral gap of the transition matrix. Let $\lambda_{\mathrm{sg}}$ be the subleading eigenvalue (or pair of complex eigenvalues) that governs the slowest relaxation mode. We define the spectral coherence
\begin{equation}
\Rcoh=\pi\frac{|\operatorname{Im}\lambda_{\mathrm{sg}}|}
{|\operatorname{Re}\lambda_{\mathrm{sg}}|},
\label{eq:coherence}
\end{equation}
which estimates the number of oscillations completed before the mode decays. How $\Rcoh$ depends on $\gex/\gin$ for fixed system size $N=3$ and various values of the dimensionality $D$ is shown in Fig.~\ref{fig:coherence}(c). We find that $\Rcoh$ is close to zero for $\gex/\gin<1$ and rises around the topological transition $\gex/\gin\approx 1$, ultimately reaching a plateau as $\gex/\gin\gg 1$. Importantly, we observe that, with increasing $D$, the transition becomes increasingly sharp and the plateau value increasingly high.

These observations can be understood by noting that, in the limit $\gex/\gin\gg 1$,  the edge cycle is dominated by the slow internal transitions, and effectively becomes a directed one-dimensional cycle with $2DN$ transitions at rate $\gin$. Thus, in this limit the expected period is $T \approx 2DN/\gin$ and the two measures of coherence coincide and are given by $\mathcal{N} \approx \Rcoh \approx 2DN $ [dotted lines in Fig.~\ref{fig:coherence}(b,c)] \cite{BaratoSeifert2017}. Thus, the topological transition converts the dynamics from incoherent high-dimensional fluctuations into a coherent unicyclic clock.

\section{A bulk topological invariant signals the transition}

The change in boundary dynamics at $\gamma_{\mathrm{ex}}=\gamma_{\mathrm{in}}$ is accompanied by a symmetry-protected bulk topological transition. Under periodic boundary conditions, Fourier transformation reduces the transition matrix to a $2D\times2D$ matrix $W(\mathbf{k})$ for each wavevector $\mathbf{k}$ (Methods). We characterize its topology using the band connected to the uniform bulk steady state, with eigenvalue $\lambda_0(\mathbf{0})=0$. Along each coordinate-axis loop $\mathbf{k}=q\mathbf{e}_i$ with $q\in[-\pi,\pi]$, this eigenvalue is real, nondegenerate, and has strictly larger real part than every other eigenvalue whenever $\gamma_{\mathrm{ex}}\neq\gamma_{\mathrm{in}}$ (Methods). Other bands may meet at exceptional points, where eigenvalues and eigenvectors coalesce, without affecting this isolated band. We normalize its right and left eigenvectors by $\langle\phi_0(\mathbf{k})|\psi_0(\mathbf{k})\rangle=1$ and define the biorthogonal Zak phase, which generalizes the usual Zak phase to non-Hermitian matrices \cite{GarrisonWright1988,Zak1989,KawabataEtAl2019,BergholtzEtAl2021}, as
\begin{equation}
Z_i=
\int_{-\pi}^{\pi}
i\langle\phi_0(q\mathbf{e}_i)|
\partial_q\psi_0(q\mathbf{e}_i)\rangle\,dq.
\label{eq:zak_band}
\end{equation}

\begin{figure}
\centering
\includegraphics[width=\linewidth]{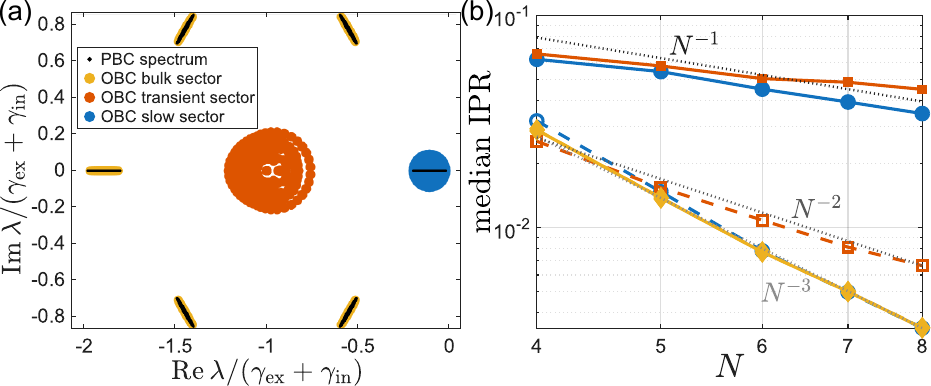}
\caption{\textbf{Localization is supported by a structured non-Hermitian skin effect.}
(a) Complex spectrum under open boundaries (OBC) compared with the periodic-boundary spectrum (PBC), for $D=3$, $N=6$, and $\gex/\gin=10$. OBC modes are separated into slow, transient, and bulk sectors by continuity with the $\gamma_{\mathrm{in}}=0$ limit (Methods). The first two concentrate on one-dimensional boundaries, while the latter stay bulk-extended.
(b) Finite-size scaling of the median right-eigenvector (solid lines) and biorthogonal (dashed lines) inverse participation ratios (IPR) for the three sectors, for $D=3$ and $\gex/\gin=10$. The color coding is as in (a). The scaling laws are shown as dotted lines. The slow sector right-eigenvector density scales as $N^{-1}$, consistent with support on the one-dimensional global cycle, whereas their biorthogonal density remains $D$-dimensional.}
\label{fig:spectrum}
\end{figure}

The transition matrix obeys a symmetry of order $2D$. Let the operation $\hat C$ shift the internal state by one step, and define $\rho(k_1,\ldots,k_D)=(-k_D,k_1,\ldots,k_{D-1})$. Then $\hat C W(\mathbf{k})\hat C^{-1}=W(\rho\mathbf{k})$. Moreover, applying the operation $D$ times reverses all wavevector components and shifts the internal state by $D$, so that $\hat I\equiv\hat C^D$ acts as inversion: $\hat I W(\mathbf{k})\hat I^{-1}=W(-\mathbf{k})$. Note that $\hat C^{2D}=\hat I^2=1$. At the inversion-invariant points $\mathbf{k}=\mathbf{0}$ and $\mathbf{k}=\pi\mathbf{e}_i$, the steady-state band is also an eigenstate of inversion, i.e.~$\hat I|\psi_0(\mathbf{k})\rangle = \xi_0(\mathbf{k})|\psi_0(\mathbf{k})\rangle$, with the inversion eigenvalue $\xi_0(\mathbf{k})\in\{+1,-1\}$. Importantly, inversion symmetry quantizes the complex biorthogonal Zak phase to $0$ or $\pi$ (modulo $2\pi$) via $e^{iZ_{i}} = \xi_0(\mathbf{0})\xi_0(\pi\mathbf{e}_i) = (-1)^{\nu_{i}}$, where $\nu_{i}\in\{0,1\}$ is the corresponding topological index \cite{TsubotaEtAl2022}.

In our system, the $2D$-fold symmetry ensures that the index is independent of the loop coordinate, $\nu_{i}=\nu$. Moreover, we can establish analytically that  $\xi_0(\mathbf{0})=+1$ and $\xi_0(\pi\mathbf{e}_i) =\operatorname{sgn}(\gamma_{\mathrm{in}}-\gamma_{\mathrm{ex}})$ (Methods), which results in
\begin{equation}
\nu
=
\begin{cases}
0, & \gamma_{\mathrm{ex}}<\gamma_{\mathrm{in}},\\[2pt]
1, & \gamma_{\mathrm{ex}}>\gamma_{\mathrm{in}} .
\end{cases}
\label{eq:bulk_z2}
\end{equation}
Equivalently, $Z_{i}=0$ below the transition and $Z_{i}=\pi$ above it. At $\gamma_{\mathrm{ex}}=\gamma_{\mathrm{in}}$, the steady-state band meets other bands and the topological index changes. The quantization is unchanged by continuous perturbations that preserve inversion and keep the steady-state band isolated along the loop.

\section{Localization is supported by a structured non-Hermitian skin effect}

The localization and funneling behavior studied above is suggestive of a non-Hermitian skin effect (NHSE), in which eigenmodes of a non-Hermitian lattice become strongly sensitive to the presence of boundaries \cite{YaoWang2018,OkumaEtAl2020,BergholtzEtAl2021}. To study this localization effect beyond the steady state, we examine the eigenvalues and eigenvectors of all $N_\mathrm{tot}=2DN^D$ modes of a finite lattice. The complex eigenvalue spectra of the transition matrix under open (OBC) and periodic (PBC) boundary conditions are shown in Fig.~\ref{fig:spectrum}(a). Deep in the topological regime $\gex\gg\gin$, the OBC spectrum separates into distinct sectors that can be connected to the exactly solvable limit $\gamma_{\mathrm{in}}=0$ (Methods). We identify a slow sector, which contains the steady state and the global-cycle oscillations; a transient sector, associated with fast relaxation through the boundary funnel; and the remaining $2D-1$ spectral sectors, which we group into a bulk class. Importantly, we observe a marked difference between the OBC and PBC spectra, a standard spectral signature of the NHSE \cite{YaoWang2018,YokomizoMurakami2019,OkumaEtAl2020}.

To quantify the NHSE, we turn to finite-size scaling of the right-eigenvector and biorthogonal densities \cite{KunstEtAl2018}. In a skin mode, the left and right eigenvectors acquire opposing localization envelopes, which cancel in their biorthogonal product. Comparing the two densities therefore distinguishes skin accumulation of the right eigenvectors from their underlying biorthogonal spatial support. The scaling of their inverse participation ratios (IPR) can further reveal the dimensionality of these supports. An example for $D=3$ is shown in Fig.~\ref{fig:spectrum}(b). Strikingly, the slow sector contains $\mathcal{O}(N^D)$ modes, yet its right density is concentrated on the one-dimensional global cycle: its median right IPR scales approximately as $N^{-1}$, while the biorthogonal IPR scales as $N^{-D}$. Thus a bulk-extensive number of modes undergoes skin localization onto a set of codimension $D-1$. The transient sector, with $\mathcal{O}(N^{D-1})$ modes, is likewise concentrated transversely onto the larger set of one-dimensional hypercube edges, whereas the remaining $\mathcal{O}(N^D)$ modes in the bulk class stay extended.

High-codimension skin localization, including corner and higher-order skin effects, is known in non-Hermitian systems \cite{KawabataSatoShiozaki2020,ZhangEtAl2021,ZhangYangFang2022}. Here, however, the dimension of the dominant right-eigenvector support and the number of localized modes strongly decouple: an extensive $\mathcal{O}(N^D)$ sector is concentrated on a one-dimensional set whose codimension $D-1$ can grow arbitrarily with $D$. Moreover, unlike canonical corner-skin examples, the high-codimension support is a system-spanning, current-carrying cycle rather than a terminal corner \cite{KawabataSatoShiozaki2020,ZhangEtAl2021,ZhangYangFang2022}. Taken together, we identify a sector-selective NHSE, in which only selected spectral sectors undergo skin accumulation, with unusually high-codimension support. This provides a spectral counterpart of the interdimensional funneling mechanism.

\section{Discussion}

The topological mechanism described here provides an extreme form of dimensional reduction: a system with a $D$-dimensional configuration space is funneled onto a single one-dimensional cycle. Counterintuitively, this reduction becomes more effective as the dimensionality of the underlying configuration space increases. High dimensionality can therefore be a resource for organizing nonequilibrium dynamics, providing both stronger confinement and more coherent cycling. This principle may be relevant beyond stochastic networks, particularly to high-dimensional and non-Hermitian quantum topological phases of matter, where our results provide an example of an extensive set of modes acquiring arbitrarily high-codimension support. In contrast to previous results on topological localization in classical stochastic systems \cite{agudo2025topological,Murugan2017,Dasbiswas2018,sawada2024role,Knebel2020,Tang2021,yoshida2022non}, the phenomena we describe have no direct counterpart in known quantum condensed matter or active matter systems.

In the biological context, our findings suggest a possible functional advantage of the repeated-subunit architecture common to protein machines. An oligomeric complex with many independently modifiable subunits possesses an enormous combinatorial state space, yet appropriately organized nonequilibrium transitions can exploit this complexity to generate a robust molecular clock: increasing the number of subunits increases its timing precision, rather than creating additional sources of disorder. In systems such as the cyanobacterial clock protein KaiC, experimental signatures of such a mechanism would include sequential modification of neighboring subunits, an increase of oscillatory coherence with oligomer size, and system-spanning modification waves whose chirality is opposite to that of the underlying local reaction cycles. More generally, our results raise the possibility that the very high-dimensional configuration spaces generated by multimerization, multisite phosphorylation, or other combinatorial post-translational modifications can be harnessed by driven biochemical networks to produce simple, precise and robust dynamical functions.

\bibliography{references}

\section*{Methods}

\subsection*{Transition matrix and steady state}

We write the master equation as
\begin{equation}
\frac{d\boldsymbol{p}}{dt}=W\boldsymbol{p}
\label{eq:master}
\end{equation}
where $W$ is the transition matrix with off-diagonal elements $W_{ab}=w_{b\rightarrow a}$ $(a\ne b)$ and diagonal elements $W_{aa}=-\sum_{b\ne a}w_{a\rightarrow b}$, where $w_{b\rightarrow a}$ is the transition rate from state $b$ to state $a$, and $\boldsymbol{p}$ is a vector containing the probabilities of all $2DN^D$ states.

For the finite open system, the steady state was computed as the normalized
right null vector of $W$:
\[
W\boldsymbol{p}^{\,\mathrm{ss}}=0,\qquad
\sum_a p^{\,\mathrm{ss}}_a=1.
\]
We used MATLAB's eig() solver to compute  both the steady-state as well as the full eigenspectrum.

\subsection*{Bulk topology}

\paragraph*{Fourier-transformed transition matrix.}
We consider $D\geq2$. Internal states are labelled
$s=0,\ldots,2D-1$ in the order
$1^+,\ldots,D^+,1^-,\ldots,D^-$, with external displacements
\begin{equation}
\mathbf{d}_s=
\begin{cases}
\mathbf{e}_{s+1}, & 0\leq s<D,\\
-\mathbf{e}_{s-D+1}, & D\leq s<2D.
\end{cases}
\end{equation}
With the Fourier convention
$\widetilde p_s(\mathbf{k})
=\sum_{\mathbf{x}}e^{i\mathbf{k}\cdot\mathbf{x}}p_{\mathbf{x},s}$
and columns labelling source states, the Fourier-transformed transition matrix is
\begin{equation}
[W(\mathbf{k})]_{s's}
=
\gex e^{i\mathbf{k}\cdot\mathbf{d}_s}\delta_{s',s+1}
+\gin\delta_{s',s-1}
-(\gex+\gin)\delta_{s's},
\label{eq:fourier_transition_matrix}
\end{equation}
where internal-state indices are understood modulo $2D$.
The cyclic shift $\hat C|s\rangle=|s+1\rangle$ obeys
$\hat C W(\mathbf{k})\hat C^{-1}=W(\rho\mathbf{k})$,
with $\rho(k_1,\ldots,k_D)=(-k_D,k_1,\ldots,k_{D-1})$.
Consequently, $\hat I=\hat C^D$ is a unitary, Hermitian inversion
operator satisfying $\hat I^2=1$.

\paragraph*{Proof that the band connected to the steady state is isolated.}
We remove the common diagonal from the transition matrix by defining the shifted matrix
$A(\mathbf{k})=W(\mathbf{k})+(\gex + \gin)\mathbb{I}_{2D}$ and also define the corresponding shifted eigenvalues 
$E=\lambda+\gex+\gin$. The characteristic polynomial of $A(\mathbf{k})$ along a coordinate-axis loop $\mathbf{k}=q\mathbf{e}_i$ can be written as
\begin{eqnarray}
\det[E\mathbb{I}_{2D}-A(q\mathbf{e}_i)]
& = &
\left[E^2-4\gex \gin \cos^2(q/2)\right]F_{D-1}(E)^2 \nonumber \\
&& -(\gex^D-\gin^D)^2.
\label{eq:axis_characteristic}
\end{eqnarray}
where we have defined the polynomials $F_m(E)$ recursively as $F_0(E)=1$, $F_1(E)=E$, and
\begin{eqnarray}
F_m(E)=E F_{m-1}(E)-\gex \gin F_{m-2}(E)
\end{eqnarray}
for $m\geq2$. Note that the roots of $F_m$ are given by
$2\sqrt{\gex\gin}\cos[j\pi/(m+1)]$ with $j=1,\ldots,m$.
Note also that the characteristic polynomial (\ref{eq:axis_characteristic}) has two terms, the second of which is always negative. As a shorthand for the following discussion, let us define
\begin{equation}
H_q(E)=\left[E^2-4\gex\gin\cos^2(q/2)\right]F_{D-1}(E)^2
\end{equation}
which corresponds to the first term. Note that all roots of $H_q$ are real, and moreover that its largest root is given by
\begin{equation}
E_{\mathrm c}(q)=2\sqrt{\gex\gin}\max\!\left\{
|\cos(q/2)|,\cos(\pi/D)\right\}.
\end{equation}
For real $E>E_{\mathrm c}$, $H_q(E)$ increases
strictly from zero to infinity. Thus, whenever $\gex\neq \gin$, there is a
unique real solution $E_0(q)>E_{\mathrm c}(q)$ of
$H_q(E_0)=(\gex^D-\gin^D)^2$.

This solution also has strictly larger real part than every other
root of Eq.~\eqref{eq:axis_characteristic}. Indeed, writing $H_q$ as a
product of its real linear factors gives
$|H_q(z)|\geq H_q(\operatorname{Re}z)$ for
$\operatorname{Re}z\geq E_0$, with a strict inequality if
$\operatorname{Im}z\neq0$. No complex root can therefore have real part
$\geq E_0$. Furthermore, $H_q'(E_0)>0$, so the eigenvalue $E_0$ is simple.
It follows that
$\lambda_0(q\mathbf{e}_i)=E_0(q)-(\gex+\gin)$ defines an isolated band on
the entire loop whenever $\gex \neq \gin$. At $q=0$, $E_0=\gex+\gin$ and $\lambda_0=0$, identifying it
as the continuation of the uniform steady state. Importantly, at $\gex=\gin$ and
$q=\pi$, the characteristic polynomial becomes
$E^2F_{D-1}(E)^2$, whose largest root
$2\gex\cos(\pi/D)$ is multiple. The separation from the other bands therefore
closes precisely at the transition value $\gex=\gin$.

\paragraph*{Inversion eigenvalues of the steady-state band.}
At $\mathbf{k}=\mathbf{0}$, the eigenvector of the steady-state band
is uniform and therefore its inversion eigenvalue is
$\xi_0(\mathbf{0})=+1$. At $\mathbf{k}=\pi\mathbf{e}_i$, the shifted
transition matrix $A(\pi\mathbf{e}_i)$ commutes with inversion. It thus
preserves the two $D$-dimensional subspaces with inversion parity
$\eta=\pm1$. Let $A_\eta(\pi\mathbf{e}_i)$ denote its restriction to
the subspace with parity $\eta$. Representing this restriction in the
basis $(|s\rangle+\eta|s+D\rangle)/\sqrt{2}$ with $s=0,\ldots,D-1$,
we obtain the parity-subspace characteristic polynomial
\begin{equation}
\det[E\mathbb{I}_D-A_\eta(\pi\mathbf{e}_i)]
=
E F_{D-1}(E)+\eta(\gex^D-\gin^D).
\label{eq:parity_characteristic}
\end{equation}
The full characteristic polynomial is the product of these two ($\eta=\pm1$)
parity-subspace polynomials.

For conciseness, we write in the following
$E_0=E_0(\pi)$ for the shifted eigenvalue of the steady-state band at
$\mathbf{k}=\pi\mathbf{e}_i$. For $\gex\neq\gin$, this eigenvalue
is nondegenerate, so its eigenvector belongs to a definite parity
subspace, with $\eta=\xi_0(\pi\mathbf{e}_i)$. Consequently, $E_0$
must be an eigenvalue of the corresponding restricted matrix
$A_{\xi_0(\pi\mathbf{e}_i)}(\pi\mathbf{e}_i)$, and hence a root of
its characteristic polynomial. Equation~\eqref{eq:parity_characteristic}
therefore gives
\begin{equation}
E_0F_{D-1}(E_0)
+\xi_0(\pi\mathbf{e}_i)(\gex^D-\gin^D)=0.
\label{eq:subspace}
\end{equation}
Because 
$E_0>2\sqrt{\gex\gin}\cos(\pi/D)$ as obtained above, we have $E_0>0$ and $F_{D-1}(E_0)>0$ and therefore $E_0 F_{D-1}(E_0)>0$. The existence of a solution of Eq.~\ref{eq:subspace} then requires that $\xi_0(\pi\mathbf{e}_i)$ and $(\gex^D-\gin^D)$ have opposite sign, so that 
\begin{equation}
\xi_0(\pi\mathbf{e}_i)
=
-\operatorname{sgn}(\gex^D-\gin^D)
=
\operatorname{sgn}(\gin-\gex).
\end{equation}
Thus the steady-state band is even under inversion for $\gin>\gex$
and odd for $\gex>\gin$.

\subsection*{Spectral sectors and non-Hermitian skin-effect}

For fully open boundaries, in real space, right and left eigenvectors were obtained
from
\begin{equation}
W\lvert\psi_n\rangle
=
\lambda_n\lvert\psi_n\rangle,
\qquad
\langle\phi_n\rvert W
=
\lambda_n\langle\phi_n\rvert,
\end{equation}
and paired by eigenvalue, with each pair normalized such that
$\langle\phi_m\vert\psi_n\rangle=\delta_{mn}$. For periodic boundaries, we sampled
wavevectors $\mathbf{k}$ uniformly in $[-\pi,\pi]^D$ and diagonalized the
corresponding $2D\times2D$ Fourier-transformed transition matrix
$W(\mathbf{k})$.

The spectral sectors in the topological regime were defined from the limit
$\gamma_{\mathrm{in}}=0$, in which every microscopic state has either one
external transition at rate $\gamma_{\mathrm{ex}}$ or no outgoing transition
because that external step is blocked by the boundary. The resulting directed graph  has three types of components. First, there are
\begin{equation}
N_\mathrm{efc}=(N-1)^D
\end{equation}
unbroken external futile cycles of length $2D$ in the bulk. The transition matrix on each such cycle is
$\gamma_{\mathrm{ex}}(P_{2D}-I_{2D})$, where $P_{2D}$ is the cyclic
permutation matrix of order $2D$, and therefore contributes the external futile cycle eigenvalues
\begin{equation}
\lambda_m^{(\mathrm{efc})}
=
\gamma_{\mathrm{ex}}
\left[
\exp\!\left(\frac{2\pi i m}{2D}\right)-1
\right],
\qquad
m=0,\ldots,2D-1,
\label{eq:strong_driving_centers}
\end{equation}
with multiplicity $N_\mathrm{efc}$ for each $m$. Note that $\lambda_0^{(\mathrm{efc})}=0$.

Second, for each internal state there are $N^{D-1}$ boundary unit cells at which
its external transition is blocked. The number of blocked microscopic states
is consequently
\begin{equation}
N_\mathrm{blocked}=2DN^{D-1},
\end{equation}
and each contributes a zero eigenvalue.

Third, all remaining microscopic states belong to directed transient chains
that terminate at blocked states, because every microscopic state has at most one incoming
and one outgoing external transition. Their number is
\begin{eqnarray}
N_\mathrm{trans} & = & N_\mathrm{tot}-2DN_\mathrm{efc}-N_\mathrm{blocked} \nonumber \\
& = &2D\left[N^D-(N-1)^D-N^{D-1}\right].
\end{eqnarray}
Ordering the transient states within
each chain by their distance from its blocked terminal state makes the
corresponding transition-matrix block triangular, with all diagonal entries
equal to $-\gamma_{\mathrm{ex}}$. Hence the eigenvalue $-\gamma_{\mathrm{ex}}$ has
algebraic multiplicity $N_\mathrm{trans}$.

As $\gin$ increases from zero, the eigenvalues form clusters that spread out from the $\gin=0$ limiting values just described. The cluster evolving from the zero eigenvalue contains those evolving from $\lambda_0^{(\mathrm{efc})}$, and those from the blocked states, for a total of
\begin{equation}
N_{\mathrm{slow}}
=
N_\mathrm{efc}+N_\mathrm{blocked}
=
(N-1)^D+2DN^{D-1}
\end{equation}
slow modes [blue in Fig.~\ref{fig:spectrum}(a)]. The cluster evolving from $-\gamma_{\mathrm{ex}}$ contains $N_{\mathrm{trans}}$ transient modes [red in Fig.~\ref{fig:spectrum}(a)]. Finally, each of the other $2D-1$ clusters contains $N_\mathrm{efc}$ modes evolving from $\lambda_{m\neq 0}^{(\mathrm{efc})}$, for a
total
\begin{equation}
N_{\mathrm{bulk}}
= (2D-1)N_\mathrm{efc}=
(2D-1)(N-1)^D.
\end{equation}
bulk modes [yellow in Fig.~\ref{fig:spectrum}(a)]. These counts satisfy
\begin{equation}
N_{\mathrm{slow}}
+
N_{\mathrm{trans}}
+
N_{\mathrm{bulk}}
=
N_\mathrm{tot}.
\end{equation}

Numerically, for the purpose of Fig.~\ref{fig:spectrum}, each eigenvalue was
assigned to one of these sectors according to its proximity to $\lambda_{m\neq 0}^{(\mathrm{efc})}$, $-\gex$, or $0$ in the complex plane. We checked that
the observed multiplicities of each cluster agreed exactly with
$N_{\mathrm{slow}}$, $N_{\mathrm{trans}}$, and $N_\mathrm{efc}$.

To determine the spatial support of each mode, internal states ($s$) were first
summed within each unit cell ($\mathbf{x}$). We defined the
normalized right density
\begin{equation}
p_n^R(\mathbf{x})
=
\frac{
\sum_s
\left|
\psi_{n,\mathbf{x}s}
\right|^2
}{
\sum_{\mathbf{x},s}
\left|
\psi_{n,\mathbf{x}s}
\right|^2
}
\end{equation}
and the normalized positive biorthogonal density
\begin{equation}
p_n^{LR}(\mathbf{x})
=
\frac{
\sum_s
\left|
\phi_{n,\mathbf{x}s}^{*}
\psi_{n,\mathbf{x}s}
\right|
}{
\sum_{\mathbf{x},s}
\left|
\phi_{n,\mathbf{x}s}^{*}
\psi_{n,\mathbf{x}s}
\right|
}.
\end{equation}
Right and biorthogonal densities distinguish anomalous accumulation of right
eigenvectors from the spatial support retained after combining the left and
right eigenvectors \cite{KunstEtAl2018}. Their inverse participation
ratios (IPR) are defined as
\begin{equation}
\mathrm{IPR}_n^R
=
\sum_{\mathbf{x}}
[p_n^R(\mathbf{x})]^2,
\qquad
\mathrm{IPR}_n^{LR}
=
\sum_{\mathbf{x}}
[p_n^{LR}(\mathbf{x})]^2.
\end{equation}
A distribution extended over a $d$-dimensional set has
$\mathrm{IPR}\sim N^{-d}$. In particular, a distribution confined
transversely to and extended along a one-dimensional path has
$\mathrm{IPR}\sim N^{-1}$.
Fig.~\ref{fig:spectrum}(b)
reports the median $\mathrm{IPR}^R$ and $\mathrm{IPR}^{LR}$ within each
sector as functions of $N$.

\setcounter{figure}{0} 
\renewcommand{\thefigure}{S\arabic{figure}}
\clearpage
\begin{figure*}[p]
\centering
\includegraphics[width=\textwidth]{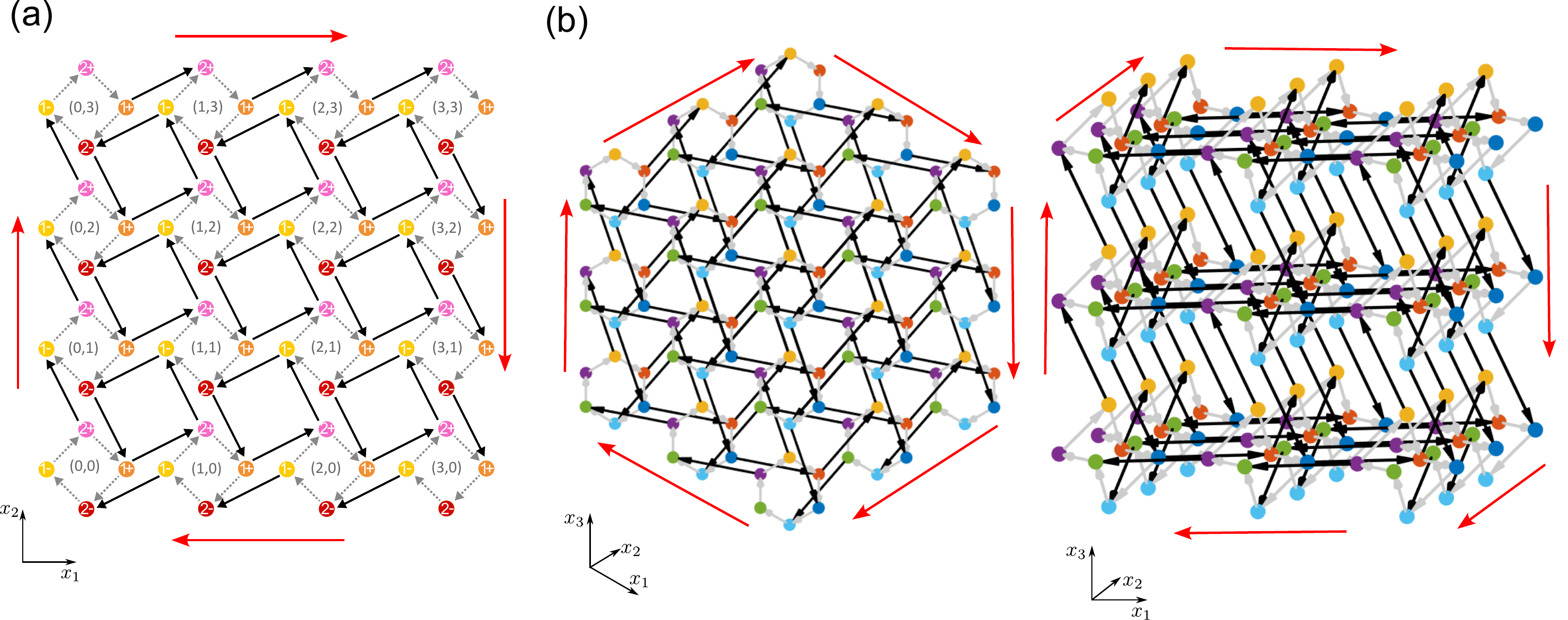}
\caption{\textbf{Full lattices in two and three dimensions.}
(a) Full lattice for $D=2$ and $N=4$. Each configuration
contains the four internal states $1+,2+,1-,2-$. Black links are external
transitions at rate $\gex$ that move between unit cells; gray links are internal
transitions at rate $\gin$ that stay within a unit cell. Red arrows trace the
global edge cycle which exists if $\gex \gg \gin$. (b) Full lattice for $D=3$ and $N=4$,
shown from two different perspectives. The global cycle (red arrows) runs along six edges of the cube.}
\label{fig:supp_lattices}
\end{figure*}

\end{document}